\documentclass[twocolumn,amsmath,amssymb,nofootinbib,tighten,floatfix,prb]{revtex4}

\usepackage{graphicx}
\usepackage{dcolumn} 
\usepackage{bm, bbm}      
\usepackage{curves}
\usepackage{epic}
\usepackage{wasysym}
\usepackage{epsf, times}
\usepackage{subfigure}
\usepackage{eufrak}
\usepackage{amsthm}

\begin{document}

\title{
Topological order and topological entropy in classical systems
      }

\author{
Claudio Castelnovo and Claudio Chamon
       }
\affiliation{
Physics Department, Boston University, Boston, MA 02215, USA
            }

\date{\today}

\begin{abstract}
We show that the concept of topological order, introduced to describe 
ordered quantum systems which cannot be classified by broken symmetries, 
also applies to classical systems. 
We discuss some of the fundamental properties of this type of 
classical order, 
and propose how to expose it via a generalized topological entropy.
Starting from a specific example, we show how to use (quantum) pure
state density matrices to construct corresponding (classical)
thermally mixed ones that retain precisely half of the original
topological entropy, a result that we generalize to a whole class of
systems.
\end{abstract}

\maketitle

\def\openone{\leavevmode\hbox{\small1\kern-4.2pt\normalsize1}}

\newcommand{\beq}{\begin{equation}}
\newcommand{\eeq}{\end{equation}}
\newcommand{\bea}{\begin{eqnarray}}
\newcommand{\eea}{\end{eqnarray}}
%
%

\section{Introduction} 

\begin{figure}[ht]
\vspace{0.2 cm}
\includegraphics[width=0.98\columnwidth]{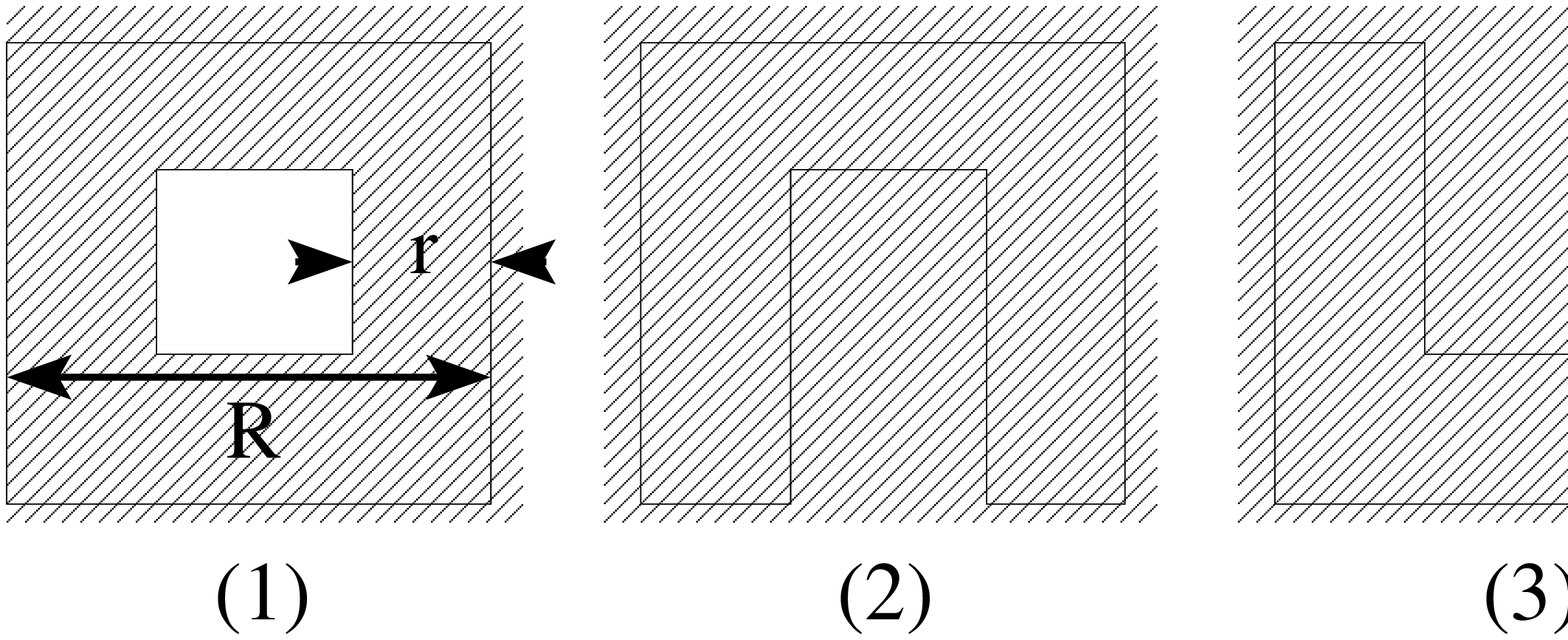}
\caption{
\label{fig: topological partitions}
Illustration of the four bipartitions used to compute the topological 
entropy in Ref.~\onlinecite{Levin2006}. 
}
\end{figure}

Over the past two decades, the notion of topological order has been developed 
to describe quantum systems exhibiting exotic properties such as a 
ground state (GS) degeneracy that cannot be lifted by any local 
perturbations~\cite{Haldane1985,Wen1990} and 
fractionalized degrees of freedom.~\cite{Arovas1984} 
These systems have been shown to exhibit a type of order that defies 
the canonical classification \textit{\`a la} Landau-Ginzburg, and that 
was named topological order.~\cite{topo refs} 
Levin and Wen, and Kitaev and 
Preskill recently proposed the idea that an appropriately defined topological 
entropy could be used to measure the presence of topological order in the GS 
of a quantum system.~\cite{Levin2006,Kitaev2006} 
Such entropy can be obtained as a linear combination of Von Neumann 
entanglement entropies of different bipartitions of the system, 
\beq
S^{\ }_{\textrm{topo}} 
= 
\lim^{\ }_{r, R \to \infty} 
\left( 
  - S^{\ }_{1A} 
  + S^{\ }_{2A} 
  + S^{\ }_{3A} 
  - S^{\ }_{4A} 
\right), 
\label{eq: old topological entropy}
\eeq
aimed at removing all bulk and surface terms to uncover the sole
topological contribution.  A particular choice of the four
bipartitions, as used in Ref.~\onlinecite{Levin2006}, is illustrated
in Fig.~\ref{fig: topological partitions}.

Despite all the efforts to understand topological order so far, the 
research focused exclusively on zero-temperature quantum systems, and the 
possibility of realizing topological order in classical systems has been 
left open.~\cite{Wen2004} 

In this paper we explicitly show that the concept of topological order
is not a pure zero temperature property but it applies also to
classical systems. Starting from a known example of quantum
topological order in a toric code, the Kitaev model,~\cite{Kitaev2003} 
we show how an appropriate coupling to a thermal bath can, in a certain
limit of the coupling constants, bring the system into a totally mixed
state that exhibits a non-vanishing topological entropy, therefore
suggesting that topological order is capable of surviving thermal
dephasing. We discuss one of the fundamental properties
of classical topologically ordered systems, namely the peculiar structure 
of phase space divided into sectors that are connected exclusively 
by extensive rearrangements of the microscopic degrees of freedom. 
These sectors are not distinguishable under any local measurement, and 
therefore lead to what ought to be called 
\emph{topological ergodicity breaking}. 
A display of glassy behavior is expected to
accompany classical topological order (and possibly {\it vice versa}). 
We briefly comment on other properties of classical topological order, 
although a thorough characterization is beyond the scope of the present 
paper. 
Finally, we reformulate the definition of topological entropy by using, 
instead of the Von Neumann entropy, a boundary entropy that is based on the 
mutual information entropy defined in information theory. 
This allows for a natural and more symmetric extension 
to classical as well as quantum systems. 
We then generalize the results on the Kitaev model to a whole class of 
quantum systems with GS wavefunctions whose amplitudes are positive and 
factorizable when expressed in terms of some (local) microscopic degrees of 
freedom. We argue that a loss of half the topological entropy between pure 
states and corresponding mixed states occurs whenever such a correspondence 
can be established. 
%
%

\section{Topological entropy in the Kitaev model} 

Consider the Kitaev model in Ref.~\onlinecite{Kitaev2003}. 
It consists of spin-$1/2$ degrees of freedom on the bonds of a 
square lattice with periodic boundary conditions. 
The Hamiltonian of the model can be written in terms of plaquette ($p$) 
and star ($s$) operators as 
\bea
H 
&=& 
- \lambda^{\ }_{A} \sum^{\ }_{s} A^{\ }_{s} 
- \lambda^{\ }_{B} \sum^{\ }_{p} B^{\ }_{p} 
\label{eq: Kitaev Hamiltonian}
\eea
where $\lambda^{\ }_{A}$, $\lambda^{\ }_{B}$ are real, positive
parameters; 
$
B^{\ }_{p} = 
\prod^{\ }_{j \in p} \sigma^{\textrm{z}}_{j} 
$ 
where $j$ labels the four spins belonging to plaquette $p$, 
and 
$
A^{\ }_{s} 
= 
\prod^{\ }_{j \in s} \sigma^{\textrm{x}}_{j}
$ 
where $j$ labels the four spins belonging to the star centered on
vertex $s$. 

Using the fact that all $A^{\ }_{s}$ and $B^{\ }_{p}$ operators have
eigenvalues $\pm 1$, that they all commute, and that they are subject
to the constraint $ \prod^{\ }_{s} A^{\ }_{s} = 1 = \prod^{\ }_{p}
B^{\ }_{p}, $ one can show that the GS of the model is degenerate and
the dimension of the GS manifold is precisely $2^{2}_{\ }$.
Furthermore, one can prove that this degeneracy is not lifted by any 
local perturbation in the thermodynamic limit, 
and the system exhibits topological order.

Let us first re-derive the topological entropy of the model with a new 
approach that can be immediately applied to both classical and quantum 
systems. 
We are going to work within one specific topological sector. 
This choice will not
affect the results as Hamma~\textit{et al.} showed that all sectors
exhibit the same entanglement entropy.~\cite{Hamma2005} 
In the $z$-basis for the spins, where all the $B^{\ }_{p}$ 
operators are diagonal, 
the GS is given by the equal amplitude
superposition of all states where the product of the four
$\sigma^{\textrm{z}}_{\ }$ components around each plaquette is
positive. Notice that the action of a star operator $A^{\ }_{s}$ in
the $z$-basis 
does not violate the plaquette constraint above.
One can show that any state in the equal amplitude
superposition (within a single topological sector) is uniquely
specified by a product $g$ of star operations required to obtain it
from a given reference state $\vert 0 \rangle$ that satisfies as well
the plaquette constraint (and lies in the same topological sector).
The GS can thus be written as 
\beq
\vert \Psi \rangle 
= 
\frac{1}{\vert G \vert^{1/2}_{\ }} 
  \sum^{\ }_{g \in G} g \vert 0 \rangle, 
\label{eq: Kitaev GS}
\eeq
where $G$ is the (Abelian) group of all possible products of star operators, 
and $\vert G \vert = 2^{n/2-1}_{\ }$ is the number of elements in the group 
($n$ being the total number of spin degrees of freedom). 
At zero temperature, the system is described by the density matrix 
\beq
\rho 
= 
\frac{1}{\vert G \vert}
\sum^{\ }_{g,g^{\prime}_{\ } \in G} 
  g \vert 0 \rangle \langle 0 \vert g^{\prime}_{\ }. 
\label{eq: rho Kitaev quantum}
\eeq

As elegantly shown in Ref.~\onlinecite{Hamma2005} by Hamma~\textit{et
al.}, the Von Neumann entropy of any bipartition $(A,B)$ of the
system prepared in such GS is given by
\bea
S^{\ }_{A} 
&=& 
-\textrm{Tr} 
  \left( 
    \rho^{\ }_{A} \log^{\ }_{2} \, \rho^{\ }_{A}
  \right) 
\nonumber \\ 
&=& 
\log^{\ }_{2} \vert G \vert - \log^{\ }_{2} (d^{\ }_{A} d^{\ }_{B}) 
\;\;\;
\left(\vphantom{\sum} 
= S^{\ }_{B} 
\right). 
\label{eq: pure state entanglement entropy}
\eea
Here $\rho^{\ }_{A}$ is obtained by taking the partial trace over subsystem 
$B$ ($\rho^{\ }_{A} = \textrm{Tr}^{\ }_{B} \rho$) and 
$d^{\ }_{A}$ ($d^{\ }_{B}$) is the number of elements in the subgroup 
$G^{\ }_{A} \subset G$ ($G^{\ }_{B} \subset G$) of transformations acting 
solely on $A$ ($B$) and leaving $B$ ($A$) unchanged. 
Moreover, Hamma and collaborators showed that whenever subsystem $A$ 
is connected, 
then 
$d^{\ }_{A} = 2^{\Sigma^{\ }_{A}}_{\ }$ and 
$d^{\ }_{B} = 2^{\Sigma^{\ }_{B}}_{\ }$, 
where $\Sigma^{\ }_{A}$ and $\Sigma^{\ }_{B}$ are the number of single 
star operators acting solely on $A$ and $B$ respectively. 

Since we are interested in computing the topological 
entropy~(\ref{eq: old topological entropy}), 
we need to extend
the above results to the case of bipartitions involving multiple disjoint
components. Let us consider first the case when the set $A$ is composed by
two disjoint regions $A^{\ }_{1}$ and $A^{\ }_{2}$, each of them 
connected, 
as in bipartition $4$ in Fig.~\ref{fig: topological partitions}. 
While all the operators in $G$ that act solely on $A$ can be
obtained as products of star operators acting solely on $A$, the converse is
no longer true for subsystem $B$. In fact, the product of \emph{all} star
operators acting on $A^{\ }_{1}$ (i.e., both those stars 
acting solely on spins in $A^{\ }_{1}$ 
\emph{and} those boundary stars acting on spins 
in $A^{\ }_{1}$ and on spins in $B$ simultaneously)
flips all $\sigma^{\textrm{z}}_{\ }$ in $A^{\ }_{1}$ two times, therefore
leaving them unchanged, while it has a non-trivial action on $B$. 
Namely, this product gives rise to an operator that flips a string of spins 
in $B$ along the boundary of $A^{\ }_{1}$. Similarly
if we consider the subsystem $A^{\ }_{2}$, and construct the product of
\emph{all} star operators acting on $A^{\ }_{2}$. The composition of these
two operators gives the product of \emph{all} star operators acting on $A$
(i.e., both those acting solely on spins in $A$ 
\emph{and} those at the boundary acting simultaneously
on $A$ and on $B$), which is equivalent to the product of all star operators
acting solely on $B$.  Therefore, out of the two additional operators, only
one of them is a new operator on $B$ that cannot be obtained as a product of
star operators acting only on $B$.  In this specific example then, $d^{\
}_{A} = 2^{\Sigma^{\ }_{A}}_{\ }$ while $d^{\ }_{B} = 2^{\Sigma^{\ }_{B} +
  1}_{\ }$.

In the generic case of 
$A=A^{\ }_{1} \cup \ldots \cup A^{\ }_{m^{\ }_{A}}$ 
and 
$B=B^{\ }_{1} \cup \ldots \cup B^{\ }_{m^{\ }_{B}}$, 
with 
$A^{\ }_{1} \ldots A^{\ }_{m^{\ }_{A}}$ 
and 
$B^{\ }_{1} \ldots B^{\ }_{m^{\ }_{B}}$ 
disjoint and connected, we obtain 
\beq
d^{\ }_{A} = 2^{\Sigma^{\ }_{A} + m^{\ }_{B} - 1}_{\ }
\qquad \;\; 
d^{\ }_{B} = 2^{\Sigma^{\ }_{B} + m^{\ }_{A} - 1}_{\ }. 
\label{eq: disconnected d_A}
\eeq

Finally, we can compute the topological entropy of the system prepared in 
its GS~(\ref{eq: Kitaev GS}) using Eq.~(\ref{eq: old topological entropy}) 
and the four different bipartitions defined in 
Fig.~\ref{fig: topological partitions}. 
{}From Eqs.~(\ref{eq: pure state entanglement entropy}) 
and~(\ref{eq: disconnected d_A}), and from the fact that 
\bea
\Sigma^{\ }_{1A} 
+ 
\Sigma^{\ }_{4A} 
&=& 
\Sigma^{\ }_{2A} 
+
\Sigma^{\ }_{3A}
\nonumber \\
m^{\ }_{1B} = m^{\ }_{4A} 
&=& 
2 
\qquad \;
(\textrm{ and all others } = 1), 
\nonumber
\eea
we obtain 
$
S^{\ }_{\textrm{topo}} 
= 
2 
= 
\log^{\ }_{2} (D^{2}_{\ })
$, 
where $D = 2$ is the so-called topological dimension of the system, 
in agreement with the result by Levin and Wen.~\cite{Levin2006} 
%
%

\section{Classical topological entropy}

Let us consider now what happens when the system in question is coupled to 
a thermal bath that allows for dephasing and thermalization 
within the lowest lying eigenstates of the plaquette operators $B^{\ }_{p}$. 
This is the case in the limit of $\lambda^{\ }_{B} \to \infty$, 
i.e., acting as a \emph{local} hard constraint, with 
$\lambda^{\ }_{A} / T \to 0$. Later on we discuss how the physics of 
this hard-constrained system is relevant to the soft case where 
the coupling constants are finite, and 
$\lambda^{\ }_{B} \gg T \gg \lambda^{\ }_{A}$. 
In the hard-constrained regime the system is described by 
a totally mixed density matrix:~\cite{footnote: SMF Kitaev} 
\bea
\rho 
&=& 
\frac{1}{\vert G \vert}
  \sum^{\ }_{g \in G} 
    g \vert 0 \rangle \langle 0 \vert g. 
\label{eq: rho Kitaev classical}
\eea
Following the same arguments as for the pure 
state~(\ref{eq: rho Kitaev quantum}) in 
Ref.~\onlinecite{Hamma2005}, we show that 
\bea
\rho^{\ }_{A} 
&=& 
\frac{d^{\ }_{B}}{\vert G \vert} 
  \sum^{\ }_{g \in G/G^{\ }_{B}} 
    g^{\ }_{A} \vert 0^{\ }_{A} \rangle 
      \langle 0^{\ }_{A} \vert g^{\ }_{A}, 
\eea
where $G/G^{\ }_{B}$ is the quotient group, and 
$g^{\ }_{A}$ and $\vert 0^{\ }_{A} \rangle$ are given by the generic 
tensor product decompositions of 
$
\vert 0 \rangle 
= 
\vert 0^{\ }_{A} \rangle \otimes \vert 0^{\ }_{B} \rangle
$ 
and 
$g = g^{\ }_{A} \otimes g^{\ }_{B}$. 
We can then show that 
\bea
\rho^{n}_{A} 
&=& 
\left(
  \frac{d^{\ }_{B}}{\vert G \vert} 
\right)^{n-1}_{\ } 
\rho^{\ }_{A}, 
\label{eq: rho square}
\eea
and use the identity 
\beq
-
\textrm{Tr} 
  \left( 
    \rho^{\ }_{A} \log^{\ }_{2} \, \rho^{\ }_{A} 
  \right)
= 
- \frac{1}{\ln 2} \; 
\lim^{\ }_{n \to 1} 
  \frac{\partial}{\partial n} 
    \left(\vphantom{\sum} 
      \textrm{Tr} \rho^{n}_{A} 
    \right), 
\label{eq: trace identity}
\eeq
to finally obtain the Von Neumann entropy 
\bea
S^{\ }_{A} 
&=& 
\log^{\ }_{2} \vert G \vert - \log^{\ }_{2} d^{\ }_{B} 
\label{eq: partial entropy 1}
\\ 
&=& S^{\textrm{(pure state)}}_{A} + \log^{\ }_{2} d^{\ }_{A} 
\;\;\; 
\neq S^{\ }_{B}. 
\label{eq: partial entropy 2}
\eea
We will comment more on this result below. 
{}From Eq.~(\ref{eq: partial entropy 2}) we can compute the topological 
entropy~(\ref{eq: old topological entropy}): 
\bea
S^{\textrm{(mixed state)}}_{\textrm{topo}} 
&=& 
S^{\textrm{(pure state)}}_{\textrm{topo}} 
+ 
\lim^{\ }_{r,R \to \infty} 
\left(
  \log^{\ }_{2} \frac{d^{\ }_{2A} d^{\ }_{3A}}
                     {d^{\ }_{1A} d^{\ }_{4A}}
\right) 
\nonumber \\
&=& 
S^{\textrm{(pure state)}}_{\textrm{topo}} 
- 1 
= 1. 
\label{eq: S_topo classical calculated}
\eea
Notice that, while the bulk and boundary contributions present in 
the classical Von Neumann entropy in Eq.~(\ref{eq: partial entropy 2}) 
precisely cancel in the expression for the topological entropy, 
the topological contributions give a non-vanishing result, 
therefore suggesting that \emph{topological order may survive thermal mixing} 
(at least in $2D$ hard constrained systems). 

Notice also that the topological entropy of the classical model
obtained upon suppressing the off-diagonal elements in the density
matrix of the Kitaev model differs from the quantum case by a factor
of $1/2$, namely $\log_2 D$ instead of $\log_2(D^{2}_{\ })$. The
physical meaning of this factor is that each of the underlying gauge
structures of the Kitaev model (the so-called electric and magnetic
loops) is independently responsible for half of the zero-temperature
topological entropy, as we recently showed in
Ref.~\onlinecite{Castelnovo2007}.

What are the physical properties of such a classical topologically
ordered system? First, it follows from the density matrix~(\ref{eq:
rho Kitaev classical}) that the expectation value for any product of
spins ($z$-component, as these are thought as classical now) is zero
unless the spins belong to a closed, non-winding loop. The system is thus a
featureless ``spin liquid'' in a classical way (much as a
paramagnet). But it does have order, topological order, in the sense
that there are different such paramagnets distinguished by the
product of spins over a winding loop on the torus. When the
constraint $\prod^{\ }_{j \in p}
\sigma^{\textrm{z}}_{j} = +1$ is fully enforced 
($\lambda^{\ }_{B}/T \to \infty$), classically changing the
topological sector requires the concomitant flipping of a number of spins
that is of the order of the linear size $L$ of the system. The
probability for such event is of order $q^L$, 
where $q^{\textrm{length}(\ell)}_{\ }$ is the independent probability of 
the combined flipping of all the spins along a loop $\ell$ 
(equivalent to a quantum tunneling event). 
This leads to the breakdown of phase space into 
disconnected topological sectors (broken ergodicity), 
and the associated time scales grow exponentially in the
size of the system, a typical signature of 
glassiness.~\cite{Cugliandolo2002} 
Notice the topological nature of such ergodicity breaking process, in that 
the disconnected sectors cannot be distinguished by any local measurements. 

Other properties observed in quantum topologically ordered systems, such as 
fractionalized excitations, lead to less intuitive classical counterparts 
to be found, for example, in the behavior of defects. 
In our classical version of the Kitaev model they appear in pairs 
connected by fluctuating strings, and they are likely to give rise 
to unusual response and relaxation processes. 
While the topological entropy vanishes identically, in the thermodynamic
limit, as soon as the constraint is 
softened,~\cite{Castelnovo2007,Alicki2007} 
the topological nature of the defects will persist at small
enough temperatures ($\lambda^{\ }_{B} \gg T \gg \lambda^{\ }_{A}$),
very much as the fractionalized excitations do in the quantum case
($\lambda^{\ }_{B}, \lambda^{\ }_{A} \gg T$). 
In fact, at such low temperatures the defects are sparse and the constraint 
is satisfied everywhere in between. Thus the defect creation, motion and 
annihilation is locally dictated by the same ``vacuum'' structure that 
would lead to topological order, were it enforced throughout the system. 
%
%

\section{Constructing a ``boundary entropy''}

{}From Eq.~(\ref{eq: partial entropy 2}), one can see that the Von Neumann 
entropy of a classical system is not a good measure of the boundary entropy 
of a bipartition of the system, because of the explicit dependence on the 
bulk entropy of one of the two partitions. 
One can obviate the problem by using the mutual information entropy of a 
bipartition, defining the boundary entropy to be 
\beq
\label{eq: boundary entropy}
S^{\ }_{\textrm{boundary}} 
= 
\frac{1}{2} 
  \left(\vphantom{\sum} 
    S^{\ }_{A} 
    + 
    S^{\ }_{B} 
    - 
    S^{\ }_{A \cup B}
  \right). 
\eeq
Clearly $S^{\ }_{\textrm{boundary}} = S^{\ }_{A} = S^{\ }_{B}$ when
the system is prepared in a pure state. In the case of a
thermally mixed state, 
all bulk entropy contributions cancel in $S^{\ }_{\textrm{boundary}}$, 
contrarily to the Von Neumann entropies $S^{\ }_{A}$ and $S^{\ }_{B}$, 
and only boundary terms are retained. 

We propose here to use $S^{\ }_{\textrm{boundary}}$ in 
Eq.~(\ref{eq: boundary entropy}) as an alternative definition of the 
Von Neumann entanglement entropy, applicable to quantum as well as classical 
systems. 
With this definition, Eqs.~(\ref{eq: pure state entanglement entropy}) 
and~(\ref{eq: partial entropy 2}) imply that the entanglement entropy stored 
in the boundary of a bipartition of a system prepared in the pure equal 
amplitude superposition state~(\ref{eq: rho Kitaev quantum}) is twice as 
large as the entropy stored in the classical 
counterpart~(\ref{eq: rho Kitaev classical}): 
\beq
S^{\textrm{(mixed state)}}_{\textrm{boundary}} 
= 
\frac{1}{2} 
  S^{\textrm{(pure state)}}_{\textrm{boundary}}. 
\label{eq: S_boundary classical vs quantum}
\eeq
At least half the entanglement entropy in the Kitaev model has 
actually a classical origin~\cite{Vedral1997}! 

Accordingly, we redefine $S^{\ }_{\textrm{topo}}$ in 
Eq.~(\ref{eq: old topological entropy}) 
using Eq.~(\ref{eq: boundary entropy}) 
\beq
\tilde{S}^{\ }_{\textrm{topo}} 
= 
\lim^{\ }_{r, R \to \infty} 
\left( 
  - S^{(1)}_{\textrm{boundary}} 
  + S^{(2)}_{\textrm{boundary}} 
  + S^{(3)}_{\textrm{boundary}} 
  - S^{(4)}_{\textrm{boundary}} 
\right). 
\label{eq: new topological entropy}
\eeq
For the Kitaev model we immediately obtain 
\beq
\tilde{S}^{\textrm{(mixed state)}}_{\textrm{topo}} 
= 
\frac{1}{2} 
  \tilde{S}^{\textrm{(pure state)}}_{\textrm{topo}}, 
\label{eq: classical vs quantum topo entropy}
\eeq
in agreement with our direct calculation in Eq.~(\ref{eq: S_topo
classical calculated}). We argue that Eq.~(\ref{eq: new topological
entropy}) is the proper reformulation of the topological entropy by
Levin and Wen in order to extend it to classical and quantum systems
alike. Together with Eq.~(\ref{eq: S_boundary classical vs quantum}),
which we show hereafter to hold for a wide class of GS wavefunctions, the 
new definition of topological entropy implies that also Eq.~(\ref{eq:
classical vs quantum topo entropy}) holds true within the same class. 
For this type of systems, at least half of the quantum
topological order has therefore a classical origin. 
(Whether part of the remaining half of the topological entropy could be 
further ascribed to classical correlations, in the sense of 
Ref.~\onlinecite{Vedral1997}, is an interesting problem.)
%
%

\section{Beyond the Kitaev model} 

All the calculations, as well as 
Eqs.~(\ref{eq: Kitaev GS}-\ref{eq: pure state entanglement entropy}, 
\ref{eq: rho Kitaev classical}-\ref{eq: partial entropy 2}), 
generalize straightforwardly to any quantum system whose 
GS is given by the equal amplitude superposition of a given 
set of states.~\cite{footnote: group condition} 

On the other hand, extending our results to generic wavefunctions 
$
\vert \Psi \rangle 
= 
\vert Z \vert^{-1/2}_{\ } 
\sum^{\ }_{g \in G} a(g) \,g \vert 0 \rangle
$ 
is highly non-trivial because $g = g^{\ }_{A} \otimes g^{\ }_{B}$ does not 
necessarily imply that $a(g) = a(g^{\ }_{A})\; a(g^{\ }_{B})$. 
Supporting results in this direction come from considering the case of 
non-negative, \emph{local} wavefunctions, as defined by the condition that 
in the continuum limit 
$
a(g) 
\to 
\mathfrak{a}(\phi) 
= 
\mathfrak{a}^{A}_{\ }(\phi^{A}_{\ }) \;
\mathfrak{a}^{\partial}_{\ }(\phi^{\partial}_{\ })\;
\mathfrak{a}^{B}_{\ }(\phi^{B}_{\ }) $, with
$\mathfrak{a}^{\partial}_{\ }(\phi^{\partial}_{\ })$ depending only on
the boundary $\partial$ between subsystems $A$ and $B$ where the two
field configurations $\phi^{A}_{\ }$ and $\phi^{B}_{\ }$ match (and
equal $\phi^{\partial}_{\ }$).~\cite{Hamma2005a} 
This is the case, for example, of scale
invariant GS wavefunctions, such as those of systems at a quantum
critical point.~\cite{Fradkin2006} 

Consider the GS wavefunction(al) in the continuum limit 
\beq
\vert \Psi \rangle 
= 
\frac{1}{\sqrt{Z}} 
  \int D\phi \, e^{-\beta E(\phi)/2}_{\ } \vert \phi \rangle, 
\eeq
where $Z = \int D\phi \, e^{-\beta E(\phi)}_{\ }$, and $E(\phi)$ satisfies 
$
E(\phi) 
= 
E^{A}_{\ }(\phi^{A}_{\ }) 
+ 
E^{B}_{\ }(\phi^{B}_{\ }) 
+ 
E^{\partial}_{\ }(\phi^{A}_{\ },\phi^{B}_{\ })
$ 
with $E^{\partial}_{\ }(\phi^{A}_{\ },\phi^{B}_{\ })$ dependent 
only on the boundary $\partial$ between subsystem $A$ and subsystem $B$, 
for any bipartition $(A,B)$. 

As discussed in Ref.~\onlinecite{Fradkin2006}, one can start from the 
corresponding pure state density matrix 
\beq
\rho 
= 
\frac{1}{Z} 
  \int D\phi D\phi^{\prime}_{\ } \, 
    e^{-\beta \left[ E(\phi) + E(\phi^{\prime}_{\ }) \right]/2}_{\ } 
      \vert \phi \rangle \langle \phi^{\prime}_{\ } \vert
\label{eq: rho SMF continuum quantum}
\eeq
and obtain the elements of the reduced density matrix 
$\rho^{\ }_{A} = \textrm{Tr}^{\ }_{B} \rho$, 
\bea
&& 
\langle \phi^{A}_{1} \vert \rho^{\ }_{A} \vert \phi^{A}_{2} \rangle 
= 
\nonumber \\ 
&&
\qquad\quad 
\frac{1}{Z} 
  \int D\phi^{B}_{\ } \, 
    e^{-\beta \left[ 
                E^{A}_{\ }(\phi^{A}_{1}) 
                + 
                E^{B}_{\ }(\phi^{B}_{\ }) 
                + 
                E^{\partial}(\phi^{A}_{1},\phi^{B}_{\ }) 
              \right]/2}_{\ } 
\nonumber \\ 
&&
\qquad\quad 
    e^{-\beta \left[ 
                E^{A}_{\ }(\phi^{A}_{2}) 
                + 
                E^{B}_{\ }(\phi^{B}_{\ }) 
                + 
                E^{\partial}_{\ }(\phi^{A}_{2},\phi^{B}_{\ }) 
              \right]/2}_{\ } . 
\eea
One can then evaluate 
\bea
&& 
\textrm{Tr} \rho^{n}_{A} 
= 
\frac{1}{Z^{n}_{\ }} 
  \int \prod^{n}_{i=1} D\phi^{\ }_{i} \, 
    e^{-\beta \left[ 
                \sum_i E^{A}_{\ }(\phi^{A}_{i}) 
                + 
                \sum_i E^{B}_{\ }(\phi^{B}_{i}) 
	      \right] 
       }
\nonumber \\ 
&& 
\qquad
    e^{-\beta \left[ 
                \sum_i E^{\partial^{\ }}_{\ }(\phi^{A}_{i},\phi^{B}_{i}) 
                + 
                \sum_i E^{\partial^{\ }}_{\ }(\phi^{A}_{i+1},\phi^{B}_{i}) 
              \right]/2}_{\ } 
      \prod^{n}_{i=1} 
        \delta (\phi^\partial_{i},\phi^\partial_{i+1}) , 
\nonumber \\ 
\label{eq: Tr (rho_A)^n quantum}
\eea
where by construction 
$\phi^\partial_{n+1} \equiv \phi^\partial_{1}$ 
and 
$\phi^{A}_{n+1} \equiv \phi^{A}_{1}$. 
The term $\prod^{n}_{i=1} \delta (\phi^\partial_{i},\phi^\partial_{i+1})$ 
is the boundary constraint arising from the product of $\rho^{\ }_{A}$ with 
itself (which enforces the boundaries to all match, because $\phi^{A}_{i}$ matches 
$\phi^{B}_{i}$, which then matches $\phi^{A}_{i+1}$, which in turn matches 
$\phi^{B}_{i+1}$, and so on).
If we factorize the integral over $\prod^{n}_{i=1} D\phi^{\ }_{i}$ as 
$\prod^{n}_{i=1} D\phi^{A}_{i} D\phi^{B}_{i} D\phi^\partial_{i}$, we can 
simplify the integrals over boundaries and the product of boundary delta 
functions to obtain 
\bea
&&
\textrm{Tr} \rho^{n}_{A} 
= 
\frac{1}{Z^{n}_{\ }} 
  \int D\phi^\partial \, 
    e^{-\beta n E^{\partial}_{\ }} 
\nonumber \\ 
&& 
\qquad
  \int^{\ }_{\phi^\partial} \prod^{n}_{i=1} D\phi^{A}_{i} \, 
  \int^{\ }_{\phi^\partial} \prod^{n}_{i=1} D\phi^{B}_{i} \, 
    e^{-\beta \left[ 
                \sum E^{A}_{\ }(\phi^{A}_{i}) 
                + 
                \sum E^{B}_{\ }(\phi^{B}_{i}) 
	      \right] 
       }, 
\nonumber \\
\eea
where we explicitly used the assumption that all the $E^{\partial^{\
}}_{\ }(\phi^{A}_{i},\phi^{B}_{i})$ terms depend solely on the
respective boundaries $\phi^\partial_{i}$.  Now that the delta
functions are no longer present, we can address the remaining
integrals as follows. The condition that the $n$ fields $\phi^{A}_{i}$
and the $n$ fields $\phi^{B}_{i}$ agree at the boundary is equivalent
to having $1$ replica free to take any value and the remaining $n-1$
to be pinned, with fixed (Dirichlet) boundary
conditions. One thus obtains~\cite{footnote:rot}
\bea
&&
\textrm{Tr} \rho^{n}_{A} 
= 
\left(
  \frac{Z^{A}_{D} Z^{B}_{D}}{Z} 
\right)^{n-1}_{\ } 
\nonumber \\ 
&& 
\qquad\quad 
\frac{1}{Z} 
  \int D\phi^\partial D{\phi}^{A}_{\ } D{\phi}^{B}_{\ }\, 
    e^{-\beta n E^{\partial}_{\ } 
       -\beta E^{A}_{\ }({\phi}^{A}_{\ })
       -\beta E^{B}_{\ }({\phi}^{B}_{\ })
       }, 
\nonumber \\ 
\eea
where 
$
Z^{A}_{D} 
= 
\int^{\ }_{D} D\phi^{A}_{\ } \, 
  e^{-\beta E^{A}_{\ }(\phi^{A}_{\ })} 
$ 
is given by the integral over subsystem $A$ with Dirichlet boundary 
conditions, and equivalently for $Z^{B}_{D}$. 

Finally, using the identity~(\ref{eq: trace identity}), 
we can compute the Von Neumann entropy 
\bea
S^{\ }_{A} 
&=& 
- 
\log^{\ }_{2} 
  \frac{Z^{A}_{D} Z^{B}_{D}}{Z} 
+ 
\frac{1}{Z} 
  \int D\phi \, \frac{\beta}{\ln 2} E^{\partial}_{\ } e^{-\beta E(\phi)} 
\nonumber \\ 
&=& 
- 
\log^{\ }_{2} 
  \frac{Z^{A}_{D} Z^{B}_{D}}{Z} 
+ 
\frac{\beta}{\ln 2} 
  \langle 
    E^{\partial}_{\ } 
  \rangle . 
\eea

The case of the totally mixed state 
\beq
\rho 
= 
\frac{1}{Z} 
  \int D\phi \, e^{-\beta E(\phi)}_{\ } 
    \vert \phi \rangle \langle \phi \vert
\label{eq: rho SMF continuum classical} 
\eeq
involves similar calculations, where the term 
$
\prod^{n}_{i=1} 
  \delta (\phi^\partial_{i},\phi^\partial_{i+1})
$ 
in Eq.~(\ref{eq: Tr (rho_A)^n quantum}) gets replaced by 
$
\prod^{n}_{i=1} 
  \delta (\phi^\partial_{i},\phi^\partial_{i+1}) \, 
    \delta (\phi^{A}_{i},\phi^{A}_{i+1})
$. 
As a result 
\bea
&&
\textrm{Tr} \rho^{n}_{A} 
= 
\left(
  \frac{Z^{B}_{D}}{Z} 
\right)^{n-1}_{\ } 
\nonumber \\ 
&& 
\qquad\quad 
\frac{1}{Z} 
  \int D\partial D\phi^{A}_{\ } D{\phi}^{B}_{\ }\, 
    e^{-\beta n E^{\partial}_{\ } 
       -\beta n E^{A}_{\ }(\phi^{A}_{\ }) 
       -\beta E^{B}_{\ }({\phi}^{B}_{\ })} , 
\nonumber \\ 
\eea
and 
\bea
S^{\ }_{A} 
&=& 
- 
\log^{\ }_{2} 
  \frac{Z^{B}_{D}}{Z} 
+ 
\frac{1}{Z} 
  \int D\phi \, 
    \frac{\beta}{\ln 2} 
          \left[ 
            E^{\partial}_{\ } 
	    + 
	    E^{A}_{\ } 
	  \right] e^{-\beta E(\phi)} 
\nonumber \\ 
&=& 
- 
\log^{\ }_{2} 
  \frac{Z^{B}_{D}}{Z} 
+ 
\frac{\beta}{\ln 2} 
  \left( \vphantom{\sum} 
    \langle E^{\partial}_{\ } \rangle 
    + 
    \langle E^{A}_{\ } \rangle 
  \right). 
\eea

Notice that, while for an equal amplitude superposition the totally mixed 
state could be reached via an appropriate high-temperature limit 
(i.e., via coupling to a 
thermal bath), 
this is no longer true for Eqs.~(\ref{eq: rho SMF continuum quantum}) 
and ~(\ref{eq: rho SMF continuum classical}), which should be interpreted 
only as a recipe to construct an associated classical system, given the 
quantum one. 

In conclusion, we obtain 
\bea
S^{\textrm{(pure state)}}_{A} 
&=& 
\beta F^{\ }_{A} 
+ 
\beta F^{\ }_{B} 
- 
\beta F^{\ }_{A \cup B} 
+ 
\frac{\beta}{\ln 2} \langle E^{\partial}_{\ } \rangle 
\nonumber \\ 
S^{\textrm{(mixed state)}}_{A} 
&=& 
\beta F^{\ }_{B} 
- 
\beta F^{\ }_{A \cup B} 
+ 
\frac{\beta}{\ln 2} \langle E^{\partial}_{\ } \rangle 
+ 
\frac{\beta}{\ln 2} \langle E^{A}_{\ } \rangle, 
\nonumber
\eea
where 
$\beta F^{\ }_{A} = - \log^{\ }_{2} Z^{A}_{D}$, 
$\beta F^{\ }_{B} = - \log^{\ }_{2} Z^{B}_{D}$, and 
$\beta F^{\ }_{A \cup B} = - \log^{\ }_{2} Z$. 

For the boundary entropy defined in Eq.~(\ref{eq: boundary entropy})
we find that
\bea
&&
S^{\textrm{(mixed state)}}_{\textrm{boundary}} 
=
\nonumber \\ 
&& 
\qquad
= 
\frac{1}{2}\left(
\beta F^{\ }_{B} 
- 
\beta F^{\ }_{A \cup B} 
+ 
\frac{\beta}{\ln 2} \langle E^{\partial}_{\ } \rangle 
+ 
\frac{\beta}{\ln 2} \langle E^{A}_{\ } \rangle
\right.
\nonumber\\
&&
\qquad\qquad 
+
\beta F^{\ }_{A} 
- 
\beta F^{\ }_{A \cup B} 
+ 
\frac{\beta}{\ln 2} \langle E^{\partial}_{\ } \rangle 
+ 
\frac{\beta}{\ln 2} \langle E^{B}_{\ } \rangle
\nonumber\\
&&
\qquad\qquad
+
\left.
\beta F^{\ }_{A \cup B} 
-
\frac{\beta}{\ln 2} \langle E^{A \cup B}_{\ } \rangle 
\right)
\nonumber\\
&&
\qquad
=
\frac{1}{2}\left(
\beta F^{\ }_{A} +\beta F^{\ }_{B} 
- 
\beta F^{\ }_{A \cup B} 
+ 
\frac{\beta}{\ln 2} \langle E^{\partial}_{\ } \rangle 
\right)
\nonumber\\
&&
\qquad
=
\frac{1}{2}\;S^{\textrm{(pure state)}}_{\textrm{boundary}} 
\;.
\eea
Thus, once again, for the boundary entropy defined in Eq.~(\ref{eq:
boundary entropy}) we find that Eq.~(\ref{eq: S_boundary classical vs
quantum}) (and therefore Eq.~(\ref{eq: classical vs quantum topo
entropy})) {\it still holds}. For any quantum system whose GS
wavefunction is \emph{local}
and with positive amplitudes, there exists an
associated classical system which exhibits precisely half the topological 
entropy. The configurations of the classical system have as 
Boltzmann weights the squares of the amplitudes in the associated
quantum GS. 
%
%

\section{Conclusions}

In this paper we show how the concept of
topological order applies to classical systems. We propose a
generalization of the definition of topological
entropy~\cite{Levin2006,Kitaev2006} that applies to classical and
quantum systems alike. We use this topological entropy to identify
topologically ordered phases at the classical level, and we discuss
some of the properties of such classical systems. We also show how to
construct thermally mixed (classical) density matrices that retain
precisely half the topological entropy of associated pure state
density matrices for a special class of quantum systems.
Our results imply that for quantum systems whose GS 
wavefunction in a given basis can be gauged so as to have non-negative 
(local) amplitudes, quantum topological order is present if and only if 
the corresponding classical system obtained by removing all off-diagonal 
matrix elements of the quantum density matrix is also topologically 
ordered as defined above. 

The result that topological order applies to classical systems could
possibly provide a way to classify classical orders without an obvious
order parameter. We end with a speculative note on the possibility
that this might be the case in glassy systems. No order parameter can
be constructed in glasses from equal-time correlation functions (as
opposed to the two-time Edwards-Anderson order parameter) of physical
degrees of freedom (as opposed to replicated variables). This fact is
suggestive that, much as in quantum topologically ordered systems like
spin liquids, no local order parameter can detect the underlying order
of the glassy states, and the hidden order could indeed be
topological. In this case, our finding that topological order and
topological entropy can be defined in classical systems may have
implications to understanding the physics of glassy systems.
%
%

\section*{Acknowledgments}

We would like to thank Xiao-Gang Wen for enlightening discussions. This
work is supported in part by the NSF Grants DMR-0305482 and
DMR-0403997 (CC and CC).
%
%


\begin{thebibliography}{99}
\bibitem{footnote:rot} 
The boundary condition of the $n-1$ pinned
replicas is determined by the field configuration (at the boundary) of
the single replica that has free boundary conditions. The fixed
boundary condition for the $n-1$ replicas should be averaged. One case
in which it is not necessary to average over boundary conditions is
that of a free bosons, in which case a rotation
%
%
$
\tilde{\phi}^{A}_{1} 
=
\frac{1}{\sqrt{n}} 
  \sum^{\ }_{i} \phi^{A}_{1} 
,\;
\tilde{\phi}^{A}_{i} 
=
\frac{\phi^{A}_{i}-\phi^{A}_{i-1}}{\sqrt{2}} 
,\;
i = 2, \dots, n. 
$
%
%
makes it clear that $\tilde{\phi}^{\ }_{1}$ is free while all others
have Dirichlet (i.e., vanishing) boundary conditions.
\end{thebibliography}
\end{document}